\documentclass[useAMS,usenatbib]{mn2e}

\usepackage{graphicx}

\newcommand{\hii}{\mbox{[H\,{\sc ii}]}}

\newcommand{\mgii}{\mbox{Mg\,{\sc ii}}}
\newcommand{\oii}{\mbox{[O\,{\sc ii}]}}
\newcommand{\oiii}{\mbox{[O\,{\sc iii}]}}
\newcommand{\neiii}{\mbox{[Ne\,{\sc iii}]}}
\newcommand{\nev}{\mbox{[Ne\,{\sc v}]}}
\newcommand{\ha}{\mbox{H$\alpha$}}
\newcommand{\hb}{\mbox{H$\beta$}}
\newcommand{\hg}{\mbox{H$\gamma$}}
\newcommand{\hd}{\mbox{H$\delta$}}
\newcommand{\W}{$W^{\lambda 2796}_0$}
\newcommand{\kms}{\mbox{km\,s$^{-1}$}}
\newcommand{\usmgii}{\mbox{USMgII}}

\title[Galactic wind hosts at $z\simeq 0.7$.]{Large scale outflows from $z \simeq 0.7$ starburst galaxies identified via ultra-strong MgII quasar absorption lines.}
\author[D. B. Nestor et al.]
{\parbox[t]{\textwidth}{\raggedright Daniel B.~Nestor$^{1,2}$\thanks{dbn@astro.ucla.edu}, Benjamin D.~Johnson$^{1}$, Vivienne Wild$^{3}$, Brice M\'{e}nard$^{4}$, David A.~Turnshek$^{5}$, Sandhya Rao$^{5}$ and Max Pettini$^{1,6}$}
\vspace*{6pt}\\
$^{1}$Institute of Astronomy, University of Cambridge, Madingley Road, Cambridge, CB3 0HA, UK\\
$^{2}$Department of Physics and Astronomy, University of California, Los Angeles, CA 90095-1547, USA\\
$^{3}$Institut d'Astrophysique de Paris, UMR 7095, 98 bis Bvd Arago, 75014 Paris, France\\
$^{4}$Canadian Institute for Theoretical Astrophysics, University of Toronto, 60 St. George Street, Toronto, Ontario, M55 3H8, Canada\\
$^{5}$Department of Physics and Astronomy, University of Pittsburgh, Pittsburgh, PA 15260, USA\\
$^{6}$International Centre for Radio Astronomy Research, University of Western Australia, 35 Stirling Highway, Crawley, WA 6009, Australia\\
}
\begin{document}

\date{}

\pagerange{\pageref{firstpage}--\pageref{lastpage}} \pubyear{2010}

\maketitle

\label{firstpage}

\begin{abstract}
Star formation-driven outflows are a critically important phenomenon in theoretical treatments of galaxy evolution, despite the limited ability of observational studies to trace galactic winds across cosmological timescales.  It has been suggested that the strongest \mgii\ absorption-line systems detected in the spectra of background quasars might arise in outflows from foreground galaxies.  If confirmed, such ``ultra-strong'' \mgii\ (\usmgii) absorbers would represent a method to identify significant numbers of galactic winds over a huge baseline in cosmic time, in a manner independent of the luminous properties of the galaxy.  To this end, we present the first detailed imaging and spectroscopic study of the fields of two \usmgii\ absorber systems culled from a statistical absorber catalog, with the goal of understanding the physical processes leading to the large velocity spreads that define such systems.  

Each field contains two bright emission-line galaxies at similar redshift ($\Delta v \la 300$ \kms) to that of the absorption.  Lower-limits on their instantaneous star formation rates (SFR) from the observed \oii\ and \hb\ line fluxes, and stellar masses from spectral template fitting indicate specific SFRs among the highest for their masses at these redshifts.  Additionally, their 4000\AA\ break and Balmer absorption strengths imply they have undergone recent ($\sim 0.01$ - 1 Gyr) starbursts.  The concomitant presence of two rare phenomena -- starbursts and \usmgii\ absorbers -- strongly implies a causal connection.  We consider these data and \usmgii\ absorbers in general in the context of various popular models, and conclude that galactic outflows are generally necessary to account for the velocity extent of the absorption.  We favour starburst driven outflows over tidally-stripped gas from a major interaction which triggered the starburst as the energy source for the majority of systems.  Finally, we discuss the implications of these results and speculate on the overall contribution of such systems to the global SFR density at $z \simeq 0.7$.
\end{abstract}


\section{Introduction}

In the local Universe, large scale gas outflows are observed to arise in galaxies exhibiting high surface densities of star formation.  While the precise roles of such outflows, including galactic ``superwinds'', in galaxy evolution are still being determined, simulations suggest that the balance between outflows and the accretion of cool gas is one of the primary mechanisms by which star formation is regulated in individual halos (e.g., Oppenheimer et al., 2009; Brooks et al., 2009).   At the current epoch, the highest star formation rate (SFR) surface densities -- and therefore galactic winds -- are preferentially found in relatively low-mass halos, such as those hosting dwarf starburst galaxies.   However, the mass of halos containing the highest specific star formation rates (sSFRs) are thought to increase with increasing look-back time, consistent with ``top-down'' galaxy formation scenarios (Cowie et al., 1996; Neistein et al., 2006).  Tracking the occurrence of galactic winds across cosmic time, particularly in manners unbiased by luminosity or halo mass, would provide a powerful way to study such models.  Furthermore, outflows are believed to be required to
explain a wide variety of astrophysical observations, from the shape of the galaxy luminosity function (Benson et al., 2003; Khochfar et al., 2007), to the stellar mass-metallicity relation (Tremonti, 2004; Erb et al., 2006; Brooks et al., 2007; Finlator \& Dav\'{e}, 2008), the large extent of dust and metals in galactic halos and in the intergalactic medium (Scannapieco, Ferrara, \& Madau, 2002; Oppenheimer \& Dav\'{e}, 2006; Kobayashi et al., 2007), and many other related phenomena.  Despite their clear importance in the galaxy formation process, outflows have generally been overlooked theoretically and, until recently, have proved difficult to study observationally at redshifts $z \simeq 1 - 4$, the epoch when the Universe formed most of its stars and superwinds were ubiquitous.

Surveys have identified outflowing gas from galaxies at redshift $z>0$ through strong, blue-shifted resonance-line absorption in their spectra arising in low-ion gas entrained in the flows (e.g., Pettini et al., 2001; Shapley et al., 2003; Tremonti, Moustakas, \& Diamond-Stanic, 2007; Wiener et al., 2009).  However, there are two important limitations inherent in methods which rely only on spectra of the outflow hosts.  First, they provide no information on the location of the outflowing gas; it is only presumed that the material reaches the IGM.  Secondly, these surveys searched for evidence of outflows in spectra of either the brightest galaxies at the relevant redshift, known star-forming galaxies, or known post-starburst galaxies.  What is needed is the reverse experiment: a survey of the galaxies from which known large-scale outflows originate.  This begs the question: how does one identify a galactic wind without {\it a priori} knowledge of the galaxy itself?  Quasar absorption lines may offer such an opportunity, as they select galaxies based on the gas absorption cross section, with no direct dependence on emission from the galaxy.

The physical processes that determine the properties of intervening low-ion quasar absorption line systems are not well understood.  While it has long been known that such absorbers can in general be identified with individual galaxies (e.g., Bergeron \& Boiss\'{e}, 1991; Steidel, Dickinson, \& Persson, 1994) correlations between emission (i.e., of the galaxy) and absorption (i.e., strength and velocity structure of the absorbing gas) properties have been elusive (Kacprzak et al., 2007) and/or inconclusive (e.g., Steidel et al., 2002; Kacprzak et al., 2010).  The
strongest absorbers, which have until recently been neglected due to their relative scarcity, may hold important clues.  For example, Bond et al.\ (2001) considered the velocity profiles of the strongest \mgii\ absorbers known at the time (rest equivalent widths \W$\sim 2$\AA) measured with high-resolution spectroscopy, and proposed that such systems may arise in galactic superwinds.

Detections of outflows through broad low-ion absorption in the spectra of starbursting galaxies suggest that galactic winds {\it can} result in very strong \mgii\ absorption along a sightline past a galaxy.  However, this does not necessarily imply that all or even any of the strongest intervening absorbers detected in quasar spectra actually {\it do} arise in the winds of foreground galaxies.  Indeed, models have been proposed that account for the observed distribution of \mgii\ absorption strength without relying on outflows (e.g., Tinker \& Chen 2008).  Alternatively, the huge kinematic spreads that define the strongest systems may be due simply to the chance intersection of the sightline with multiple ``normal'' \mgii\ absorbing galaxies in a rich group or cluster, as first suggested by Pettini et al.\ (1983).

Very large \mgii\ absorber surveys (e.g., Nestor et al., 2005; Prochter, Prochaska, \& Burles, 2006; Quider et al., 2010), which are now becoming available, have uncovered large numbers (e.g., $>600$ in Quider et al.) of ``ultra-strong'' \mgii\ (\usmgii) systems with \W$\ge 3$\AA.  Catalogs of such systems afford the opportunity to explore in depth the proposed \usmgii\ absorber-galactic wind connection.  Recent work has already given support for a connection between the strongest \mgii\ systems and star forming galaxies, which is usually considered as support of an outflow scenario. For example, by stacking thousands of relatively-shallow Sloan Digital Sky Survey (SDSS) images of the fields of strong \mgii\ absorption systems Zibetti et al.\ (2007) demonstrated the strongest systems are associated with bluer galaxies closer to the sightline to the background quasar compared to weaker systems.  Similarly,  Bouch\'{e} et al.\ (2007) have detected strong H$\alpha$ emission at the absorption redshift towards strong \mgii\ absorbers and Rubin et al.\ (2009) have identified an \usmgii\ absorber in the spectrum of a background galaxy that they identify with a wind from a foreground galaxy.  Perhaps the most compelling evidence suggesting a connection between \mgii\ absorbers and star formation is the relation between \W\ and \oii\ emission discussed by M\'{e}nard et al.\ (2009), wherein they demonstrate that the strongest \mgii\ absorbers are on average associated with the highest \oii\ luminosity densities and are therefore likely to be hosted by vigorously star-forming galaxies.

Nestor et al.\ (2007; hereafter NTRQ) published the first imaging survey aimed specifically at the strongest \mgii\ absorption systems, including images of the fields of thirteen moderate-redshift ($0.42 < z < 0.84$) \usmgii\ systems.  These revealed bright galaxies at relatively low impact parameter to the absorption sightline (compared to the fields of most \mgii\ absorbers).  While consistent with the outflow model in general and, e.g., the results of Zibetti et al., in particular, detailed study of the galaxies associated with \usmgii\ absorbers is needed to test this putative connection.  If the ``ultra-strong'' nature of these absorbers is indeed linked to galactic winds, we expect to find evidence of recent, high mass fraction starbursts in one or more of the low impact parameter (low-$b$) galaxies.  In this paper, we present the results of an imaging and spectroscopic study of the galaxies detected in two \usmgii\ absorber fields from the NTRQ sample, conducted to test both the \usmgii\ absorber/outflow connection and alternative models.  In \S2 we describe the selection of targets, observations and reductions of our data, and our basic observational results.  In \S3 we discuss the properties of the galaxies determined to be at similar redshift to the \usmgii\ systems.  We present further discussion in \S4 and \S5, and summarize the paper in \S6.  Throughout the paper we assume a cosmology with $\Omega_M = 0.3$, $\Omega_\Lambda = 0.7$ and $H_0 = 70$~{\mbox{km\,s$^{-1}$\,Mpc$^{-1}$} and state magnitudes in the AB system.

\section{Data}
\subsection{Target Selection}
\label{Sec:targets}
The targets for imaging in NTRQ were chosen from the now publicly-available Pittsburgh \mgii\ absorber catalog
(Quider et al., 2010) based solely on \W, absorption redshift and observability, and thus should represent an otherwise
unbiased sample of \usmgii\ absorber environments at those redshifts.  
Many of the images from that study reveal a fairly bright ($L \ga 0.3 L^*$) isolated galaxy or clump of emission at small
impact parameter ($b \la 3$\arcsec) to the sightline, representing clear candidates for the object(s) associated with the absorption.  

We were awarded 23 hours of the Gemini-North queue in semester 2008A
to obtain spectra of these candidates using the Gemini Multi-Object Spectrograph (GMOS).
However, as the time was awarded in ``band 3'', it was likely that the observations would take
place in sub-optimal conditions.  Thus, it would be exceedingly difficult to obtain observations of
targets with particularly faint apparent magnitudes or those with small angular separation from the background quasar.
Therefore, we limited the observations to long exposures of three fields from NTRQ: 
those towards SDSS J074707.62+305415 (hereafter Q0747+305; $z_{abs} = 0.765$, \W$=3.63$\AA), 
SDSS J101142.01+445155.4 (Q1011+445; $z_{abs} = 0.836$, \W$=4.94$\AA) and 
SDSS J141751.84+011556.1 (Q1417+011; $z_{abs} = 0.669$, \W$=5.6$\AA).  
The primary absorber-galaxy candidates in these fields
are all relatively bright ($m_i < 21.9$) -- although the long exposures
allow the determination of redshifts for fainter targets, as well -- 
and at sufficiently large impact parameter ($b > 4$\arcsec) to place
slitlets that avoid the point spread function of the quasar even in poor seeing.  
These are the three fields that NTRQ categorize as ``Bright'' in their descriptions of the
\usmgii\ absorber environments.  It is important to emphasize that, as these fields differ
in appearance from the majority of those imaged in NTRQ, they might not be
representative of \usmgii\ systems in general.  We discuss this further in \S\ref{sect:Discusion}.

We show the \mgii\ absorption region of the SDSS spectrum of the quasar in the Q0747+305 and Q1417+011 fields in figures \ref{Fig:0747} and \ref{Fig:1417}.  The absorber towards Q0747+305 has an observed profile consistent with the minimum possible velocity spread given its \W, $\Delta v^{rest} \simeq 390$~\kms.  For the absorber towards Q1417+011, the doublet members are blended but the profiles appear to span $\Delta v \approx 1000$~\kms.  The vertical lines in each figure correspond to relative velocity offsets of galaxies in the field (see below). 

\subsection{Observations and reductions}
Pre-imaging was obtained for the three fields with the $r$-band filter on 
GMOS early in semester 2008A.  As 
the WIYN photometry from NTRQ utilized the $i$-band for these three fields, 
the relatively deep pre-imaging
also provided $r-i$ ($\sim$~rest-frame $U-B$ at $z\sim 0.7$) colours for sources detected in both
data sets.  Reduced and combined images were provided by
Gemini.  SExtractor (Bertin \& Arnouts, 1996) was used for deblending of sources and obtaining 
photometric measurements.  Photometric zero-points were determined using bright unsaturated 
sources in our images having SDSS magnitudes available from the SDSS website.

Slit masks were designed with 2\arcsec-wide slitlets
and lengths which varied to optimize placement in the crowded regions near the quasar
while still allowing for individual sky measurements.  
Spectra were obtained in the Q0747+305 field on 4 and 13 March 2008 with total 
integration time of 22800s over twelve individual exposures, and in the Q1417+011 field on 1, 2, 10
and 12 May and 11 June 2008 with a total integration time of 28500s over fifteen exposures.
A single exposure of 1900s in the Q1011+445 field was obtained on 12 April 2008, but with an insufficient
signal to noise ratio for scientific purposes; we thus limit further discussion to the Q0747+305 and Q1417+001
fields.

All spectra were obtained using the R400 grating and OG515 filter with three different grating
tilts to obtain continuous coverage across the detector gaps.  Spectra were reduced and 
extracted in standard fashion with exposures of quartz halogen and CuAr lamps used
for flat-fielding and wavelength calibration.  Individual sky-subtracted extracted spectra were
combined with weighting by signal to noise ratio.  Slitlets were placed over stars in each field,
which aided in the removal of telluric features in the spectra.  A single observation of HZ44 
on 13 March 2008 was provided to facilitate relative spectro-photometric calibration.  Absolute
spectro-photometric calibration was achieved by comparing synthetic magnitudes derived
from the spectra to our photometry.

\subsection{Results}
The pre-imaging was accomplished under better than expected conditions, with low
atmospheric attenuation and seeing of $ \simeq 0.6\arcsec$ for each field.  
The images are deeper than the $i$-band images from NTRQ, with $\sigma_m \la 0.1$ down to $i \simeq 26$ for 
compact sources.

The signal to noise ratios of the resulting spectra varied with both wavelength and
the brightness of the source, and for our primary targets ranged from $\sim 5$ to $\sim 60$.
Subtraction of sky emission lines was problematic, particularly in the red wavelengths and for the 
fainter sources.  Similarly, correcting for telluric absorption was only moderately
successful.  We tested our relative spectrophotometric calibration by comparing synthetic
colours derived from the calibrated spectra of our brighter sources to colours from the SDSS, 
as well as a direct comparison of the spectrum of one source that was spectroscopically observed
by both us and the SDSS.  Various issues complicated such comparisons, such as slit losses and 
photometric uncertainties.  Thus, we are only able to put a limit on our relative spectrophotometric
accuracy of better than $\approx$ twenty per cent.  Emission lines were fit with single Gaussian profiles (no emission multiplets were resolved) and line fluxes and flux uncertainties determined using an optimal extraction procedure.

\subsubsection{Q0747+305 ($z_{qso}=0.974$, $z_{abs} = 0.7646$, \W$=3.6$\AA)}
The region of the GMOS $r$-band
image surrounding the sightline for this field is shown with slitlet positions overlaid in figure~\ref{Fig:0747}.  
We were able to place slitlets over all resolved sources brighter than $i = 23.83$ within an impact
parameter of $b=70$\,kpc at $z=z_{abs}$, and brighter than $i=22.68$ within $b=200$\,kpc.  
Comparing to the characteristic magnitude $M^*_i$ at 
redshifts $0.45 < z < 0.85$ (Gabasch et al., 2006), these limits correspond to approximately
0.08 $L^*_i$ and 0.23 $L^*_i$, respectively, for a Sc-like $k$-correction.  In table~1, we present the photometry
for all galaxies detected with $m_r < 24$ within $b<200$\,kpc of the sightline at $z=z_{abs}$ and/or
with a slit spectrum from our observations, together with the determined redshift if 
spectroscopically observed.  Values of $m_i$ are from NTRQ.

\begin{table}
 \begin{centering}
  \caption{Observed galaxies towards Q0747+305 ($z_{abs} = 0.7646$)}
  \begin{tabular}{ccccccc}
    \hline
$\Delta \alpha^{\rm a}$ & $\Delta \delta^{\rm a}$ & slit no. & $b$ & $z$ & $m_r$ & $m_i$ \\
\multicolumn{2}{c}{(arcsec)}& & (kpc) & & & \\
    \hline
$-1.8$ & 4.5 & 3 & 34 & $0.339$ & 20.09 & 19.53 \\
{\bf 3.9 } & {\bf 2.9} & {\bf 2} & {\bf 36} & {\bf 0.7660} & {\bf 22.46} & {\bf 21.86}  \\
$-7.5$ & $-1.8$ & 5 & 57$^{\rm \dag}$ & ? & 23.73 & 22.99\\
{\bf 6.6} & {\bf 5.0} & {\bf 1} & {\bf 61} & {\bf 0.7643} & {\bf 21.66} & {\bf 20.67} \\
6.4 & $-9.4$ & ... & 84$^{\rm \dag}$ & ... & 23.14 & 22.68 \\
11.0 & 7.7 & 4 & 97 & $0.7215$ & 23.56 & 23.19\\
9.2 & 10.8 & ... & 105$^{\rm \dag}$ & ... &  22.90 & 22.73 \\
$-12.5$ & 7.1 & ... & 106$^{\rm \dag}$ & ... & 23.88 & 23.98 \\
$-13.1$ & 11.4 & ... & 128$^{\rm \dag}$ & ... & 23.99 & 23.65 \\
$-22.4$ & $-9.5$ & ... & 180$^{\rm \dag}$ & ... & 23.81 & 22.72 \\
16.4 & 18.3 & ... & 181$^{\rm \dag}$ & ... & 23.28 & 23.99 \\
$-11.7$ & 23.4 & ... & 193$^{\rm \dag}$ & ... & 23.69 & ... \\
24.7 & 13.6 & 6 & 218 & $0.8813$ & 23.59 & 22.94\\
31.0 & 16.6 & 7 & 256 & $0.7328$ & 21.38 & 20.46\\
$-25.6$ & 34.4 & 8 & 315 & $0.7519$ & 22.21 & 21.46\\
16.6 & 36.5 & 9 & 205 & $0.3702$ & 20.05 & 19.54\\
$-15.1$ & 49.2 & 10 & 338 & ?0.577 & 22.36 & 21.27\\
\hline
\end{tabular}
\end{centering}
Galaxies with relative velocities differences $\delta v < 350$\,km\,s$^{-1}$ to the absorption are bolded.  a: relative to quasar sightline; \dag: if $z_{gal} = z_{abs}$. \hfill
\end{table}

The spectrum of the bright low-$b$ (slit 3; 5\arcsec to the north-west of the sightline) 
galaxy in this field exhibited no emission lines.  However, identification of the absorption features
\hb,  \mbox{Mg\,b}, \mbox{Fe5270}, \mbox{Fe5406} and \mbox{Na\,D} allow the determination of a redshift $z=0.339$, i.e., in the foreground of the absorber.  The two galaxies with $b=61.3$\,kpc (slit 1; hereafter \mbox{G07-1}) and
$b=36.1$\,kpc (slit 2; \mbox{G07-2}) to the north-east 
of the sightline both exhibit strong emission lines of \oii, \oiii, \hb\ and \hg\ at redshifts similar (within $\approx 50$ - 250 \kms) to that of the absorber; see figure \ref{Fig:0747}.  Their spectra are shown in figure~\ref{Fig:0747specs}.  We also detect \neiii\,$\lambda3869$ (6828\AA\ observed) and tentatively identify \nev\,$\lambda3426$ (6045\AA\ observed) in the spectrum of \mbox{G07-1}.  The flux-calibrated spectra were used to compute rest-frame $u$-, $g$-, and $B$-band absolute magnitudes and emission-line fluxes, which we present in table~2.  We quote observed values (i.e., no attenuation correction has been applied), although \hb\ has been corrected for stellar absorption. 

We note that the quasar spectrum also exhibits \mgii\ absorption at $z=0.7219$ with 
\W$=0.5$\AA.  The compact galaxy under slit 4 has a redshift corresponding to a velocity
difference of 122 \kms\ from that absorber and $b=97$\,kpc from the sightline.

\begin{table}
  \begin{centering}
  \caption{Measured Galaxy Properties}
  \begin{tabular}{rcccc}
    \hline
    & G07-1 & G07-2 & G14-1 & G14-2\\
    \hline
\multicolumn{5}{c}{magnitudes and colors}\\
    \hline
$M_B$ & $-22.03$ & $-20.85$ & $-21.86$ & $-20.06$ \\ 
$M_g$ & $-22.42$ & $-21.39$ & $-22.09$ & $-20.25$ \\
$M_u-M_g$ & $1.49$ & $1.70$ & 0.83 & 0.59 \\
    \hline
\multicolumn{5}{c}{line fluxes ($\times 10^{17}$ ergs s$^{-1}$ cm$^{-2}$)}\\
    \hline
    \oii\ $\lambda3728$ & $30.7 \pm 0.3$ & $13.3 \pm 0.3$ & $58.2 \pm 0.3$ & $18.6 \pm 0.2$ \\
    \hg$^{\rm a}$ & $\sim 1$ & $1.5 \pm 0.2$ & $9.6 \pm 0.3$ & $\sim 1$ \\
    \hb & $6.5 \pm 0.2$ & $4.9 \pm 0.3$ & $29.0 \pm 0.2$ & $5.7 \pm 0.1$ \\
    \oiii\ $\lambda4960$ & $22.9 \pm 0.4$ & $5.4 \pm 0.6$ & $8.8 \pm 0.3$ & $3.2 \pm 0.1$ \\
    \oiii\ $\lambda5008$ & $80.4 \pm 0.9$ & $4.5 \pm 0.5$ &  $25.1 \pm 0.4$ & $13.4 \pm 0.3$\\
\hline
\end{tabular}
  \end{centering}
a: errors do not include potentially large systematic uncertainties due to telluric absorption. \hfill
\end{table}

\subsubsection{Q1417+011 ($z_{qso}=1.727$, $z_{abs} = 0.669$, \W$=5.6$\AA)}
Figure~\ref{Fig:1417} shows the GMOS $r$-band
image of this field, also with slitlet positions overlaid.  
Slitlets were placed over all resolved sources brighter than $i = 23.26$ within an impact
parameter of $b=70$\,kpc at $z=z_{abs}$ and brighter than $i=20.78$ within $b=200$\,kpc,
corresponding to approximately 0.09 and 0.84 $L^*_i$.  
Table~3 presents the photometry
and the determined redshift for the sources spectroscopically observed.

\begin{table}
  \begin{centering}
  \caption{Observed galaxies towards Q1417+011 ($z_{abs} \simeq 0.669$)}
  \begin{tabular}{ccccccc}
    \hline
$\Delta \alpha^{\rm a}$ & $\Delta \delta^{\rm a}$ & slit no. & $b$ & $z$ & $m_r$ & $m_i$ \\
\multicolumn{2}{c}{(arcsec)}& & (kpc) & & & \\
    \hline
{\bf --2.5} & {\bf --3.2} & {\bf 2} & {\bf 29} & {\bf 0.6671} & {\bf 22.67} & {\bf 21.86} \\
4.8 & $-5.5$ & ... & 51$^{\rm \dag}$ & ... & 23.54 & ... \\
$-6.1$ & $-4.7$ & ... & 54$^{\rm \dag}$ & ... & 23.30 & 23.26 \\
7.4 & 2.1 & ... & 54$^{\rm \dag}$ & ... & 23.85 & ... \\
{\bf 6.5} & {\bf 5.1} & {\bf 1} & {\bf 58} & {\bf 0.6678} & {\bf 21.04} & {\bf 20.33} \\
10.4 & $-2.0$ & ... & 74$^{\rm \dag}$ & ... & 23.08 & 23.01 \\
6.8 & $-9.5$ & 3 & 61 & 0.385 & 19.84 & 19.32\\
$-14.3$ & $-9.3$ & ... & 120$^{\rm \dag}$ & ... & 23.78 & ... \\
18.1 & $-0.9$ & ... & 127$^{\rm \dag}$ & ... & 21.58 & 21.15 \\
$-11.6$ & 14.1 & ... & 128$^{\rm \dag}$ & ... & 23.28 & 22.34 \\
$-22.6$ & 1.8 & ... & 159$^{\rm \dag}$ & ... & 22.33 & 21.27 \\
$-20.0$ & $-11.5$ & ... & 162$^{\rm \dag}$ & ... & 21.40 & 20.78 \\
$-9.4$ & $-22.0$ & ... & 168$^{\rm \dag}$ & ... & 22.38 & 21.90 \\
8.9 & 23.1 & ... & 174$^{\rm \dag}$ & ... & 23.12 & ... \\
1.3 & $-28.8$ & 5 & 207 & 0.7072 & 21.68 & 21.04\\
28.7 & 10.7 & 4 & 184 & ?0.517 & 22.18 & 21.37\\
\hline
\end{tabular}
\end{centering}
See notes to table~1.
\end{table}

As in the Q0747+305 field, the brightest low-$b$ source (slit 3; 12\arcsec to the south-east) 
lacks emission lines, but we were able to determine a redshift of $z=0.385$ from absorption features in the spectrum, placing the 
galaxy in the foreground of the absorber.  Also similar to the previous field, the two next-brightest 
galaxies in the field (hereafter G14-1 and G14-2) exhibit strong \oii, \oiii, \hb\ and \hg\ emission lines with redshifts similar (within $\approx 200$ - 350 \kms; see figure \ref{Fig:1417}) to  $z_{abs}$.  Both may also exhibit \neiii\,$\lambda3869$.  The impact parameters are $b=58.3$\,kpc and $b=28.7$\,kpc, respectively.  Their spectra are displayed in figure~\ref{Fig:1417specs} and the absolute magnitudes and line-fluxes presented in table~2.

\begin{figure*}
\begin{minipage}[c]{0.95\textwidth}
 \centering
 \includegraphics[angle=0,width=0.9\textwidth]{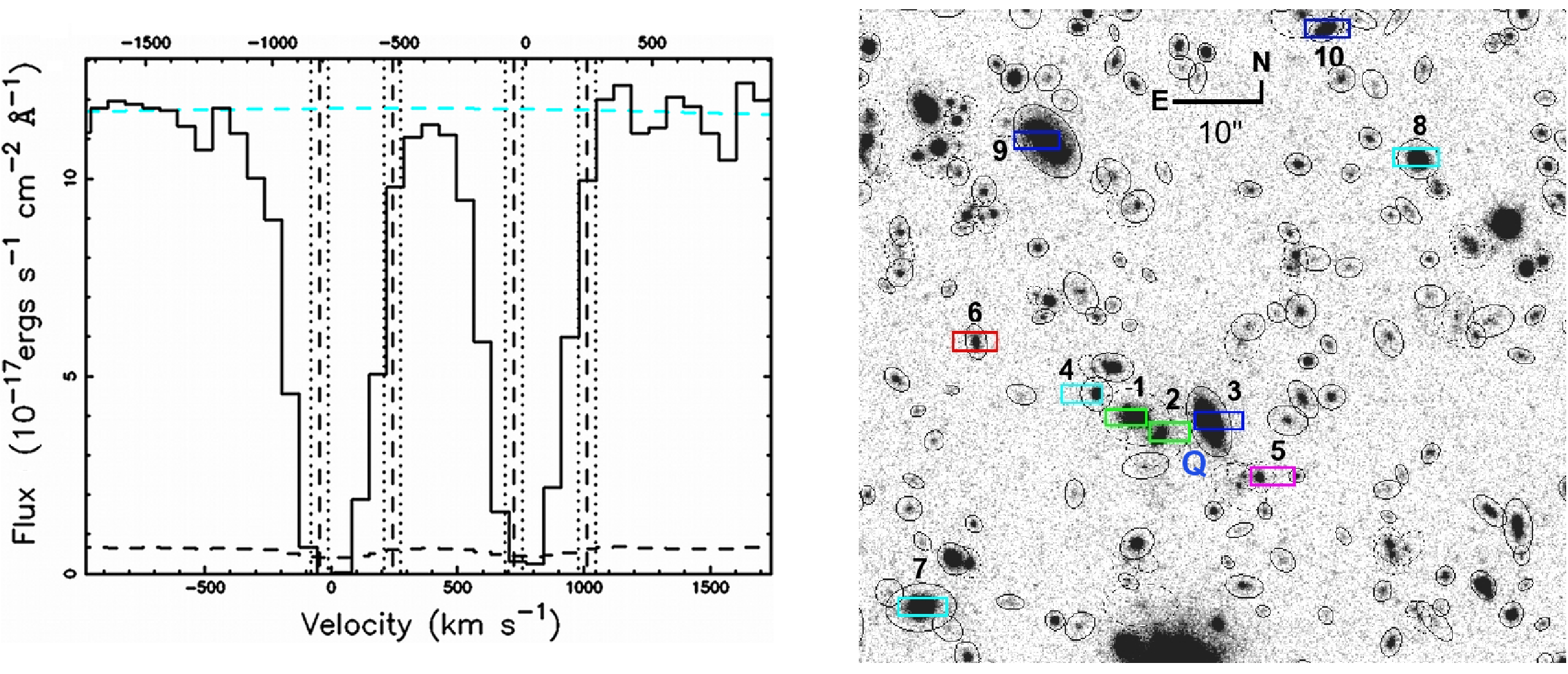}
 \caption{Left: SDSS spectrum of SDSS J074707.62+305415 highlighting the region
    of the \usmgii\ absorption doublet, with our fit to the continuum shown as the dashed curve.  The bottom axis shows rest-frame velocity relative
    to the $\lambda2796$ line; the top axis shows velocity relative to the $\lambda2803$ line.
    Vertical dashed-lines correspond to $2796 \times (1+z_{gal})$ and $2804 \times (1+z_{gal})$ for the two 
    $z_{gal} \simeq z_{abs}$ sources, with the vertical dotted-lines indicating the 
    1\,$\sigma$ uncertainty.  
    Right: GMOS $r$-band image of the Q0747+305 field.  The quasar
    PSF has been fit and subtracted, the residuals masked and the location marked 
    with a 'Q'.  The location of the slitlets are indicated and colour-coded with 
    green corresponding to sources with redshifts within a few hundred \kms\ of 
    the \usmgii\ absorption redshift, 
    cyan within a few thousand \kms, blue for foreground galaxies,
    red for background galaxies, and purple unknown redshifts.}
\label{Fig:0747}

 \includegraphics[angle=-90,width=0.9\textwidth]{fig2.ps}
  \caption{Spectra of $z_{gal} \simeq z_{abs}$ galaxies in the field towards Q0747+305.  
    Top: spectrum of \mbox{G07-1}.  Rest-frame wavelength is indicated
    on the top axis and relevant emission and absorption lines are marked.  Poor night-sky subtraction is responsible for the strong residuals.  Bottom: same, for \mbox{G07-2}.}
\label{Fig:0747specs}
\end{minipage}
\end{figure*}

\begin{figure*}
\begin{minipage}[c]{0.95\textwidth}
  \centering
 \includegraphics[angle=0,width=0.9\textwidth]{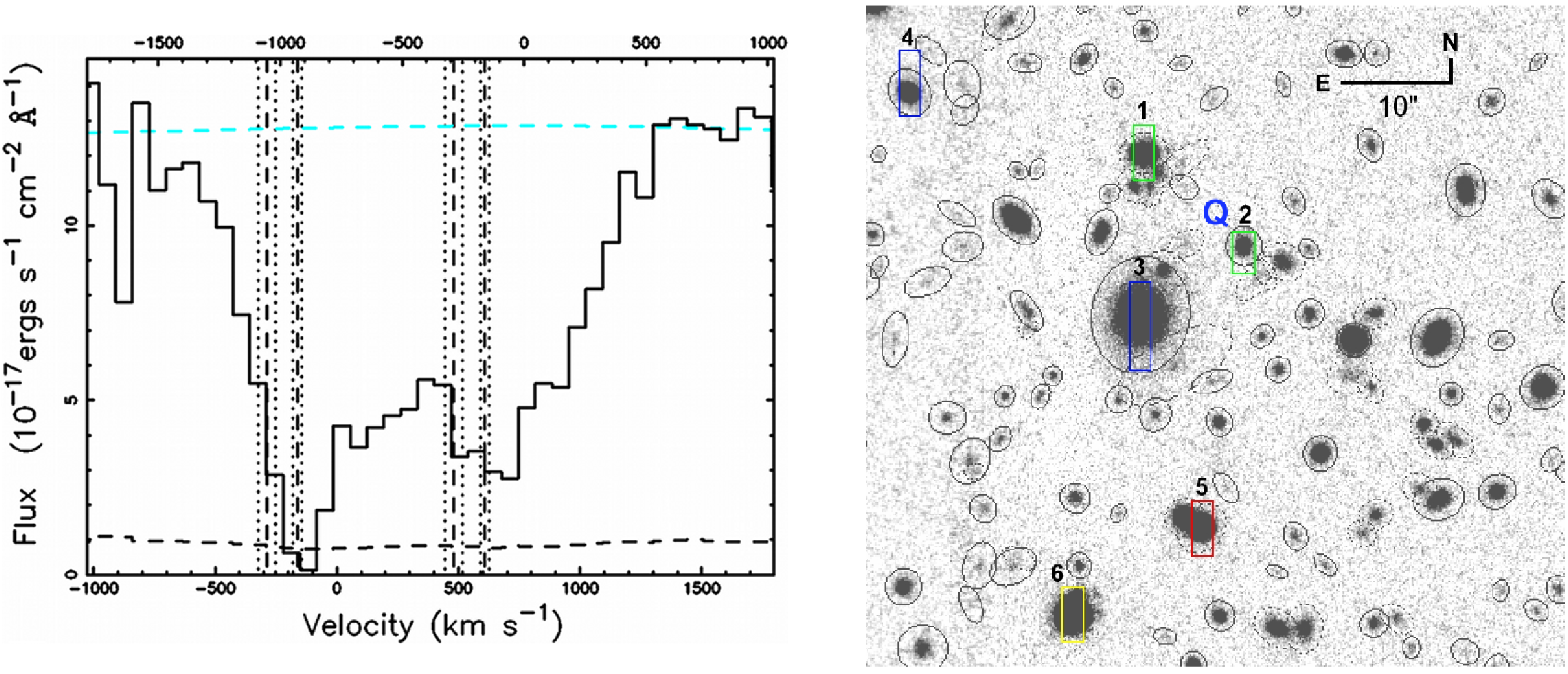}
 \caption{Same as figure 1, but for SDSS J141751.84+011556.1.  Yellow slitlet indicates a star.}
\label{Fig:1417}

  \includegraphics[angle=-90,width=0.9\textwidth]{fig4.ps}
  \caption{Same as figure 2, but for SDSS J141751.84+011556.1.  Top: \mbox{G14-1}.  Bottom: \mbox{G14-2}.}
 \label{Fig:1417specs}
\end{minipage}
\end{figure*}

\section{Properties of the $z \simeq z_{abs}$ galaxies}
In this section we detail our analysis of the spectra presented in figures \ref{Fig:0747specs} and \ref{Fig:1417specs}.  To investigate the nature of the $z \simeq z_{abs}$ galaxies, we wish to put them in the context of galaxies at similar redshift as regards their colours, metallicities, SFRs, stellar mass and starburst properties.  We use emission line measures to determine metallicities and SFRs, fit the stellar continua using a template-fitting method to explore stellar masses, and examine the 4000\AA\ break and Balmer absorption features to explore the ages and relative size of possible recent star formation episodes.  We described the computation of absolute magnitudes and colours above (\S2).  Here, we briefly detail our other methods, followed by a summary of our findings.

\subsection{Emission lines measurements}
We used the fluxes presented in table 2 to examine the instantaneous star formation rates and metallicities for each galaxy.  The location of red-shifted \hg\ coincided with regions of telluric absorption, particularly in \mbox{G07-1} and \mbox{G07-2}; the quoted uncertainties do not include the potentially large systematic errors due to our limited ability to correct for atmospheric features.  

SFRs can be approximated from our data using either \hb\ or \oii$\lambda\lambda3727,3730$ based estimates.  While \hb\ provides a more direct estimate, the blended \oii\ doublet is measured at higher significance than \hb\ in our spectra; therefore explore both methods.   For \hb-based estimates, we assume a \ha\ to \hb\ ratio of 2.86 appropriate for case B recombination (Brocklehurst, 1971) and the relation between \ha\ luminosity and SFR from Kennicutt (1998).  For \oii\ based estimates, we employ the relation between \oii\ luminosity and SFR from Kewley, Geller and Jansen (2004).  Our results are summarised in table~4.  For \mbox{G14-1} and \mbox{G14-2}, we report SFRs based on $L$(\hb).  For \mbox{G07-1} and \mbox{G07-2} the \hb\ errors may be underestimated due to the emission line partly falling on the GMOS chip-gaps in some of the exposures.  Thus, we report values derived from $L$(\oii) for these galaxies.  The two methods produced consistent (within $1 \sigma$) values for three of the four galaxies, with the exception being \mbox{G14-1}, for which \hb\ gave a higher value (12.2 versus 7.4~M$_{\sun}$~yr$^{-1}$) at $1.3 \sigma$.  

The potential for dust-induced reddening means that the SFR values should be considered lower-limits.  In principle, the level of attenuation could be modelled using the flux ratio of \hb\ to \hg.  However, large relative statistical and potential systematic (see above) errors made such a correction difficult, prohibitively so in \mbox{G07-1} and \mbox{G14-2}.  The measured \hb/\hg\ flux ratios suggest that the unreddened SFRs are larger by a factor of $\sim$ 1.2 to 2.2 in \mbox{G14-1} and by a factor of $\sim$ 1.5 to 22 in \mbox{G07-2}.  Alternatively, we also consider the average luminosity-dependent reddening corrections for star forming galaxies as described by Moustakas, Kennicutt \& Tremonti (2006), for both \hb- and \oii-derived SFRs.  As our uncorrected \hb- and \oii-derived SFRs are similar in each galaxy, the larger \oii-based average correction implies a larger true SFR than does the \hb-based average correction.  We report in table~4 the range in SFRs corresponding to that spanned by the two average corrections.

To determine galaxy metallicities we use the well-known strong-line ``$R_{23}$''-ratio method.  We employ the approach detailed in Kewley \& Dopita (2002) which iteratively estimates [O/H] and the ionization parameter to fine-tune the estimate of [O/H].  For a given value of $R_{23}$, [O/H] is double-valued.  The lower-metallicity results, however, are inconsistent with expectations from the mass-metallicity relation (Tremonti et al., 2004) given our mass estimates (see below).  Thus, we adopt values using the upper-branch of the $R_{23}$ curve.  For G07-1, we were unable to use the iterative method as our measured value of $R_{23}$ is unphysically high -- suggesting that the value we determine for \hb\ in that spectrum may indeed be inaccurate.  We thus report for G07-1 a value of [O/H] from $R_{23}$ and \oiii/\oii\ using the prescription of McGaugh (1991), though we caution that this may not be accurate.  
 
Finally, we note that the possible detections of \nev\ in the spectrum of \mbox{G07-1} indicates a possibility that an active nucleus is contributing to the observed line fluxes.  We consider typical ranges of \oii/\neiii\ (e.g., Netzer, 1990) for Seyfert 1 galaxies to estimate the possible AGN contamination, and conclude our SFR estimate may be overestimated by 10 to 70~per~cent in \mbox{G07-1}.  This possible contamination is small compared to the uncertainty due to the unknown dust correction, and in any case does not significantly affect our conclusions.

\subsection{Template fitting}
The manifestation of star formation-driven outflows depends on galaxy mass as well as the rate of star formation.  Therefore, the star formation rate per unit stellar mass -- i.e., the specific star formation rate (sSFR) -- of our galaxies is of particular interest.  The rest-frame near-IR emission in most galaxies is dominated by older stellar populations, which presumably account for the bulk of the stellar mass in galaxies at this epoch.  While we lack near-IR photometry for these galaxies, robust stellar mass estimates can also be obtained though the fitting of theoretical galaxy spectral templates to observed spectra.  Such methods are most robust when sufficient color and emission-line information is available, which ease the degeneracies in M/L due to bursts of star formation.  Thus we employ a spectral template-fitting algorithm being developed by BJ to constrain the stellar masses of each of the four $z\simeq z_{abs}$ galaxies.  The method will be described in detail in a future publication; here, we provide only a brief overview of the method.

We first convolved each spectrum with 25 non-overlapping artificial filters with FWHM$=$40-60\AA\ over the rest-frame wavelength range $~$[3525\AA,4940\AA], designed to exclude \oii$\lambda3727$ and the \hb, \hg, and \hd\ emission lines.  We then built a library of model fluxes in these artificial filters using the population synthesis models of Bruzual \& Charlot (2003), incorporating the dust attenuation model of Charlot \& Fall (2000).  To incorporate uncertainties in the calibration of the data and the systematic uncertainties of the models, we added a 0.05 mag systematic uncertainty to the resulting synthetic photometry in each band.  The parameters that determine the model fluxes are drawn from prior distributions motivated by ``typical'' libraries (e.g. Salim et al. 2007), which are in turn motivated by observational evidence when possible.  The prior parameter distributions include: (i) a flat metallicity distribution from Z=0.04 to 2; (ii) an e-folding time of the exponentially decaying SFR given by a flat distribution in its reciprocal over the range 0 to 1 Gyr$^{-1}$; (iii) ages nearly-uniformly distributed from 0.1 to 6 Gyr; (iv) dust attenuation of young stars (ages $< 10^7$ yrs, i.e., \hii\ regions) given by $\tau_V$ with a $\sim$flat prior probability distribution (see da Cunha 2008); (v) dust attenuation of older stars given by $\mu \times \tau_V$ with the $\mu$ prior given by a Gaussian centered at 0.3 with 0.2 dispersion, over the range $\mu=0.1$ to 1 (Calzetti, Kinney, \& Storchi-Bergmann, 1994; Charlot \& Fall, 2000); (vi) bursts of duration 100 Myr; (vii) burst frequency 0.5 per Gyr; (viii) burst amplitudes (i.e., total mass formed in the burst relative to the ``host'' stellar mass at the time of the burst) distributed logarithmically from 0.03 to 4. The parameters drawn from these prior distributions are then rounded to the nearest value in a fine grid of parameter space for which synthetic spectra have been created. The \hb\ luminosity of each model spectrum is determined from the number of ionizing photons in the model assuming $\log L_{\mbox{H}\beta}=\log N_{ion}-12.33$ in units of erg s$^{-1}$, and that the line is attenuated with an optical depth $\tau_{\mbox{\hb}}=\tau_V(4681/5500)^{-0.7}$ (Charlot \& Fall, 2000).  In total, we use 38000 models which we can compare to the synthetic photometry and uncertainty, as well as to the measured value of \hb\ from our spectra.  The fitting procedure and construction of cumulative probability distribution functions follows Salim et al. (2007).  

The stellar masses reported in table~4 from our template fitting are the medians of the resulting cumulative probability distribution functions, with errors corresponding to the 2.5 and 97.5 percentiles of the distribution.  The best fit chi-square values per degree of freedom for our template fitting were $\sim$1 for all four galaxies.  

\subsection{4000\AA\ break and Balmer absorption strengths}

The strengths of the 4000\AA\ break and Balmer absorption lines in
galaxy spectra indicate the relative mass fractions of young,
intermediate and old stellar populations. Very young stellar
populations, dominated by hot O and B stars with short main-sequence
lifetimes, have weak 4000\AA\ break strengths and weak Balmer
lines. Intermediate age populations, dominated by A and F stars with
main sequence lifetimes of order 0.5\,Gyr, show strong Balmer
absorption lines.  Old stellar populations are dominated by the light
from G stars and cooler, with lifetimes $\ga 1$\,Gyr and have
strong 4000\AA\ break strengths. Therefore, the 4000\AA\ break region of the optical
spectrum places tight constraints on the {\it recent} star formation
of galaxies, rather than the {\it instantaneous} star formation rate
measured by nebular emission lines.

In order to characterize the strengths of these features, we fit
components (eigenspectra) derived from a principal component analysis
of model stellar populations by Wild et al.\ (2007). These
eigenspectra, created for the purpose of analysing SDSS spectra, have
been applied directly to the GMOS spectra without any alteration given
the similar spectral resolution\footnote{Note that the SDSS components
of Wild et al. (2007) are different from the components presented in
Wild et al. (2009) for the analysis of much lower resolution spectra
and can therefore not be directly compared}. The first component is
equivalent to the 4000\AA\ break strength, traditionally quantified
via the D4000, or D$_n$4000, indices. The second component measures
the deficit (excess) of Balmer absorption lines seen in starburst
(post-starburst) galaxies. The advantage of the components over the
traditional indices based upon H$\delta$ or H$\beta$ equivalent widths
alone is that they combine several Balmer lines, together with
information from the shape of the blue/UV continuum, to boost
considerably the SNR of the measurement.  In a similar way to
D4000/H$\delta$, the component amplitudes (PC1, PC2 given in the final two
columns of table~4) allow the identification of a
galaxy as belonging to the red sequence, blue cloud (star-forming),
star-bursting, or post-starburst classes of galaxies.  See, for
example, figure~9 in Wild et al.\ (2007).

Through comparison to updated spectral synthesis models of Bruzual \&
Charlot (2003), the indices can be used to constrain the age and mass
fraction of a recent episode of sharply enhanced star formation.
The method is described in detail in Wild, Heckman \& Charlot (2010).  

\begin{table*}
  \centering
  \begin{minipage}[c]{0.95\textwidth}
    \caption{Computed properties}
    \begin{tabular}{lcccccccccc}
\hline
 & SFR$_{\rm min}$ & SFR$_{\rm cor}$ & 12+log(O/H) & M$_*$ & M$_*$/L$_B$ &sSFR$_{\rm min}$ & sSFR$_{\rm cor}$ & burst age & PC1 & PC2\\
 & \multicolumn{2}{c}{(M$_{\sun}$~yr$^{-1}$)}  & & (log[M$_{\sun}$]) & & \multicolumn{2}{c}{(log[Gyr$^{-1}$])} & (Gyrs) & & \\
\hline
G07-1 & $\ga 5.4$ & 16 - 59 & 9.0$^\dag$ & $11.23^{+0.32}_{-0.12}$ & 1.8 & $> -1.5$ & $-1.0$ - $-0.5$ & $\sim 1.0$ & $-$3.61 & 0.75 \\
G07-2 & $\ga 2.4$ &  9 - 16  & 8.9 & $10.78^{+0.24}_{-0.10}$ & 1.9 & $> -1.4$ & $-0.8$ - $-0.6$ & $\sim 0.9$ & $-$3.83 & 0.77  \\
G14-1 & $\ga 12.2$ & 49 - 81 & 9.0 & $10.34^{+0.24}_{-0.16}$ & 0.3 & $> -0.25$ & 0.4 - 0.6 & $\sim 0.02$ & $-$5.17 & $-$1.29 \\
G14-2 & $\ga 2.4$ & 6 - 11 & 8.8 & $9.30^{+0.32}_{-0.04}$ & 0.1 & $> 0.1$ & 0.5 - 0.7 & $\la 0.01$ & $-$6.10 & $-$2.11 \\
      \hline
    \end{tabular}
$\dag$ may be inaccurate due to poorly-measure \hb\ flux; see text.
\end{minipage}
\end{table*}

\subsection{Estimated galaxy properties}
\label{Sec:props}
The $z\simeq z_{abs}$ galaxies are bright, with rest-frame $g$-band luminosities 2.4 $L^*$ and 0.9 $L^*$ for G07-1 and G07-2 respectively, and 1.8 $L^*$ and 0.3 $L^*$ for G14-1 and G14-2, where $L^*$ is given at similar redshift by Gabasch et al.\ (2004).  Comparing the rest-frame $u-g$ colours to local galaxies in the SDSS from Blanton et al.\ (2003)\footnote{Although the magnitudes in Blanton et al.\ are $k$-corrected to $z=0.1$, the effect on the colours is negligible.}, we find that G07-1 and G07-2 are typical of bright galaxies, while G14-1 and G14-2 are considerably bluer than most bright galaxies, suggesting SEDs dominated by young OB stars.

While we lack accurate dust-corrected SFRs estimates, all four of the galaxies clearly exhibit significant ongoing star formation.  In order to compare our estimated SFRs with those of other galaxies at similar redshifts, we consider the SFRs corresponding to the characteristic UV and H$\alpha$ luminosities (i.e., $L^*_{UV}$ and $L^*_{H\alpha}$) using luminosity functions at $z \simeq 0.7$ from the compilation of Hopkins et al.\ (2004).  We find values for ``SFR$^*$'' between 1.6 - 9.3~M$_{\sun}$~yr$^{-1}$.  Thus, galaxies with comparable or greater SFRs are relatively rare by number but account for a significant amount of the star formation density in the Universe at $z \simeq 0.7$.

The results of our template-fitting suggest that G07-1 is quite massive, sitting above the knee in the galaxy stellar mass function at $z\approx 0.7$ (e.g., Drory et al., 2009), that G07-2 and G14-1 lie in the relatively flat part of the stellar mass function, and that G14-2 is a notably low-mass galaxy.  Combining the mass estimates with the photometry, we compute B-band mass-to-light ratios which we compare to models for bursty spiral galaxies by Bell \& de Jong (2001).  G07-1 and G07-2 have M/L$_{\rm B}$ values as expected of galaxies with neutral B$-$R colours, while G14-1 and G14-2 have particularly low M/L$_{\rm B}$ values even for very blue galaxies.  Again, this implies that G14-1 and G14-2 are undergoing notably high mass-fraction starbursts.  We can also obtain estimates of the rotation velocities, $v_{rot}$, through the Tully-Fisher relation using either the absolute B magnitudes or stellar masses (e.g., Fern{\'a}ndez Lorenzo et al.\ 2009; Kassin et al.\ 2007), which, in turn, allow us to approximate the respective escape velocities, $v_{esc} \simeq 3 \times v_{rot}$ (Veilleux, Cecil, Bland-Hawthorn, 2005).  We find $v_{esc} \sim 400$-800\,\kms\ and $\sim300$-600\,\kms\ for G07-1 and G07-2, respectively, and $v_{esc} \sim 200$-800\,\kms\ and $\sim 100$-500\,\kms\ for G14-1 and G14-2, respectively.

The mass estimates allow us to compute sSFRs for each galaxy, which are shown in table~4.   These can be compared to galaxies at similar mass and redshift (Feulner et al., 2005).  The sSFRs for all four galaxies are among the highest at their mass/redshift (see Feulner et al., figure 1).  G14-1 and G14-2 are well above and G07-1 and G07-2 near the ``doubling'' line in sSFR, which is often used to distinguish between passively star forming and starbursting galaxies.  

The metallicities of each of the four galaxies are typical for their mass at $z\sim 0.7$ (Savaglio et al., 2005; Maiolino et al., 2008).  It is not clear how to simultaneously interpret  gas-phase metallicities in \hii\ region determined from galactic emission lines together with those determined from absorbing regions at significant galactocentric distance.  It is noteworthy, however, that while measurements reported in the literature for ``typical'' low-ion absorption systems are largely metal poor (e.g., Pettini et al., 1999; Meiring et al., 2008; Nestor et al., 2008), it has been demonstrated that gas-phase metallicity is correlated with \mgii\ REW (e.g., Nestor et al., 2003; Turnshek et al., 2005; Murphy et al., 2007), and those in \usmgii\ are expected to approach those measured in our galaxies.

Finally, the PC1 and PC2 amplitudes for each galaxy can be compared to the categories defined by these indices in figure~9 in Wild et al.  The points in that figure correspond to spectra of local ($0.01 < z < 0.07$) SDSS galaxies, which are dominated by light from the inner $\simeq 4$\,kpc as the SDSS fibers are generally significantly smaller than the sizes of galaxies at those redshifts.
Nonetheless, the physical interpretation of the principal component amplitudes (i.e., for categorizing stellar populations as quiescent, starburst, etc.)
is robust.  \mbox{G07-1} and \mbox{G07-2} have PC1 and PC2 amplitudes placing them firmly in the 
post-starburst region of the PC1-PC2 plane, while \mbox{G14-1} and \mbox{G14-2} are firmly in the starburst region.  Notably, these results are consistent with the findings from each of our other methods described above.  Comparing these PC1 and PC2 values to the models (\S3.3), we find estimates for the ages for the starbursts of $\approx1$~Gyr for \mbox{G07-1} and \mbox{G07-2}, $\approx20$~Myr for \mbox{G14-1} and $\la 20$~Myr for \mbox{G14-2}.  It is interesting that the star formation outbursts for the two galaxies in each field appear to be coeval.  This strongly suggests that the starbursts are related to their proximity; i.e., triggered by an interaction.

To summarise our observational results: (i) each field contains a pair of emission-line galaxies with $z_{gal} \simeq z_{abs}$, at impact parameters $b \simeq 30$\,kpc and $b \simeq 60$\,kpc to the absorption sightlines; (ii) \mbox{G07-1} and \mbox{G07-2} appear to be fairly massive, bright galaxies; (iii) \mbox{G14-1} and \mbox{G14-2} are also bright but less massive and have very blue rest-frame $u-b$ colours; (iv) all four galaxies have high sSFRs and have metallicities typical for their mass; (v) \mbox{G07-1} and \mbox{G07-2} are roughly 1 Gyr removed from a starburst phase, though both exhibit ongoing star formation; and (vi) \mbox{G14-1} and \mbox{G14-2} appear to currently be undergoing a starburst episode.  

\section{Discussion}
\label{sect:Discusion}
We undertook the observations discussed above with the goal of uncovering the physical mechanism driving the extreme absorption velocity spreads that define \usmgii\ systems.  In this section, we test the star formation-driven galactic wind model in light of the results of these observations, and find abundant circumstantial evidence in its favour.   We conclude the section by considering other popular models, and find tidally-stripped gas from interacting galaxies is also consistent with the observational results for some \usmgii\ systems.

\subsection{Evidence from Bayesian probability}
First, we consider the possibility that the presence of \mbox{(post-)}starburst galaxies in the vicinities of ultra-strong \mgii\ absorption are coincidental.  Indeed, the existence of a strong \mgii\ absorber requires a galaxy at $z \simeq z_{abs}$.  However, galaxies with such properties as we find for \mbox{G07-1}, \mbox{G07-2}, \mbox{G14-1} and \mbox{G14-2} are uncommon (e.g., Bell \& de Jong 2001; Feulner et al., 2005; Wild et al., 2009).  Similarly, the presence of a galaxy produces a likelihood of detecting a \mgii\ absorber with $z \simeq z_{gal}$, but \mgii\ systems with \W $\ge 3.6$\AA\ account for only 1 per cent of strong (\W $\ge 0.3$\AA) systems, and those with \W $\ge 5.6$\AA\ only 0.05 per cent.  Thus it is exceedingly unlikely that the (post-)starburst nature of the galaxies and the ultra-strong nature of the \mgii\ absorption are unrelated.  This does not necessarily imply the starbursts are driving a wind that is responsible for the \usmgii\ absorption.  For example, it is conceivable that they share a common cause, such as a major interaction stripping gas out to large galactic radii while simultaneously triggering a starburst, or that such an interaction channeled gas to the galaxy centers and fed galactic nuclear activity, which in turn drove the outflow.  

\subsection{Evidence from kinematic absorption spread}
\label{Sec:kin}
The minimum rest-frame velocity width of a  completely saturated, opaque \mgii\ $\lambda2796$ feature is given by $\Delta v_{min}= (\Delta\lambda/\lambda)\times c = $(\W$/$1\AA)$\times 107$\,\kms.  For \usmgii\ systems (i.e., 3\AA\ $<$ \W\ $\la$ 6\AA), this corresponds to 320 \kms\ $\la \Delta v_{min} \la$ 640 \kms.  Actual kinematic spreads are typically larger due to partially non-opaque profiles.  As discussed in \S\ref{Sec:targets}, the absorber towards Q0747+305 (figure~\ref{Fig:0747}) and Q1417+011 (figure~\ref{Fig:1417}) have observed profiles consistent with $\Delta v \simeq 390$~\kms, and $\Delta v \approx 1000$~\kms, respectively.  Thus, the physical mechanism behind \usmgii\ systems must be one that is able to produce cool gas that, along a single line-of-sight, continuously spans many hundreds of \kms\ with a total dynamic spread of up to $\ga$~1000~\kms.  Such a dynamic range can naturally be obtained along a sightline passing though a galactic wind at some angle to the outflow orientation.

\subsection{Timescale, distance, and velocity consistency}
The impact parameters of the four $z_{gal} \simeq z_{abs}$ galaxies in the present study are comparable to those of the general population of \mgii\ absorbers (see, e.g., Kacprzak et al., 2007).  They are, however, at the extreme end of the $b$-distribution for \usmgii\ systems (NTRQ).  It is therefore worthwhile to consider the inferred velocities and timescales of the putative outflows in light of the distances needed to be traversed by outflowing material to cover the sightline to the QSO.  

The observed absorption velocity spread arises from projections of the outflow velocity (of sufficient columns of low-ion gas) onto the sightline, and thus depends on the unknown geometry.  However, order of magnitude estimates of the outflow speeds can be made by considering the red- and blue-most velocity extents relative to the galaxy systemic velocity (e.g., figures \ref{Fig:0747} and \ref{Fig:1417}) together with a conic outflow geometry featuring opening angles between $\simeq 45^{\deg}$ and $100^{\deg}$ (Veilleux, Cecil, Bland-Hawthorn, 2005).  While such a calculation is overly simplistic, it should give an order of magnitude estimate for the outflow speeds which can be interpreted together with our estimations of the age of the most recent starburst.  

For the Q0747+305 field, we estimate an outflow velocity of $v \sim 300$ - 600~\kms.  This range is comparable to our approximation of the escape velocities for these galaxies (\S \ref{Sec:props}).  If outflows from both galaxies contribute to the observed profile, then the velocity could be as low as $v \sim 200$~\kms.  At these velocities, the gas would have to have flowed for a minimum of $\sim100$ - 500~Myrs to have reached the distance to the sightline.  These timescale estimates are well below our estimate of a $\sim 1$~Gyr age for the bursts in \mbox{G07-1} and \mbox{G07-2}.  Thus, the relatively large impact parameters in this field are completely consistent with a scenario involving outflows driven by the starburst event.  

The starbursts in the Q1417+011 field are likely much younger.  As can be seen in figure \ref{Fig:1417}, however, much
larger velocities are also necessary to explain the absorption.  For an outflow to account for the absorption, we estimate $v \sim 800$ - 1000~\kms, which exceeds our approximations of the escape velocities.  The velocity estimate changes little if both galaxies contribute, as they are at similar redshift and both $z<z_{abs}$.  Though rather large, such velocities are indeed seen in some galactic winds, especially at high redshift (e.g., Pettini et al., 2002; Quider et al., 2009; Dessauges-Zavadsky et al., 2010).  This translates into minimum travel times of $\sim 20$ - 200~Myrs, which is roughly consistent with our age estimates for the bursts in this field. 

\subsection{Related evidence}
Unfortunately we do not cover \mgii\ or other strong low-ion absorption features in our spectra of the $z \simeq z_{abs}$ galaxies.  Thus, we are not able to directly identify cold gas entrained in outflows via profiles blue-shifted with respect to the emission lines.  However, we reiterate that strong, broad, blue-shifted low-ion absorption is ubiquitous in the spectra of star-forming galaxies at all redshift (Heckman, 2000; Tremonti, Moustakas, \& Diamond-Stanic, 2007; Weiner et al., 2009; Vanzella et al., 2009; Steidel et al., 2010).  An example of blue shifted {\it ultra}-strong low-ion absorption in a galaxy spectrum is the well-known $z=2.7$ starbursting galaxy 1512-cB58, with REWs in \mbox{O\,{\sc i}} and \mbox{Si\,{\sc ii}} lines of 4.4\AA\ (Pettini et al., 2000).  These features are generally considered signs of large-scale outflows.  

Finally, we note that the incidence of \usmgii\ systems (Nestor et al., 2005; Nestor et al., in preparation) and the global SFR density (Hopkins et al., 2006) show a remarkably similar fractional decrease from $z \simeq 1.2$ to $z \simeq 0.7$.  The cause and consequences of the equivalence of these trends are unclear; however it suggests that \usmgii\ absorbers have some strong dependence on star formation.  

\subsection{Other possibilities}
For the \usmgii\ systems studied here, the associated galaxies have properties indicating significant recent/ongoing star formation episodes, suggesting a connection between the absorbers and star formation-driven galactic outflows.  This is in contrast to the galaxies associated with the majority of weaker \mgii\ absorption, which are found to span a range of colours and morphologies, with most considered ``normal'' galaxies.  The physical nature of these absorbers is still an open question.  Traditionally, the absorption kinematics have been thought to arise from a variety of causes including rotating gaseous disks, virialised clouds, and accreting gas.  As we have not {\it directly} observed a galactic wind in our systems, we must consider if the phenomena thought to be responsible for the absorption kinematics in weaker systems, or perhaps some other uncommon phenomenon, can lead to \usmgii\ absorption alone, without recourse to a galactic wind.

Detailed studies of disk-galaxy/\mgii-absorber pairs have shown that extended rotating gas disks likely account for part of the absorption kinematics in some systems (Steidel et al., 2002; Kacprzak et al., 2010).  However, the projected spread in velocity of a sightline through a gas disk should be less than the disk rotation velocity.  Spreads in \usmgii\ systems are much larger than those expected for galaxies of the luminosities of our galaxies (Catinella, Giovanelli, \& Haynes, 2006).  

Comparison of the neutral hydrogen gas density of galaxies with that of the cosmic star formation rate density reveals the necessity for continual replenishment of \mbox{[H\,{\sc i}]} (Hopkins McClure-Griffiths \& Gaensler, 2008).  Could the observed velocity spreads in \usmgii\ absorbers be due to sightlines through such condensing gas, e.g., in streams or clouds of accreting cool gas?  Again, it is difficult to reconcile the huge range in absorption velocity along a single sightline with accreting gas alone.  A combination of a contribution from a rotating disk and accretion is more palatable for less-ultra strong systems such as Q0747+035.  The accretion scenario, however, may favor finding the host galaxy in a pre-starburst or early starburst phase, whereas \mbox{G07-1} and \mbox{G07-2} appear to be in a post-starburst phase.

The large spread in absorption velocity may be due to the chance intersection of the sightline with multiple halos hosting less-strong ``typical'' \mgii\ absorbers.  This scenario has additional appeal for the two sightlines in the present study since they both contain a pair of galaxies at $z \simeq z_{abs},$ as opposed to the majority of the NTRQ sample which have only a single absorber-galaxy candidate.   This is particularly true for Q0747+035, in which the two galaxies bracket the absorption redshift.  This scenario would be supported if the kinematics split into two or more distinct groupings of velocity components when observed with higher resolution.  Unfortunately, we do not have such observations for our systems.  Notwithstanding, insight can be gained by considering high-spectral resolution observations of other \mgii\ absorbers, including \usmgii\ systems.  In figure~\ref{Fig:UVESspecs} we show the $\lambda2796$ or $\lambda2803$ (as the red-most components of the $\lambda2796$ lines overlap the saturated regions of the $\lambda2803$ line in two systems) transition for five $z\simeq 2$ \usmgii\ systems observed with UVES on the VLT.  These data will be discussed in detail in an forthcoming paper.  The REW of systems (c), (d) and (e) in figure~\ref{Fig:UVESspecs} are dominated by a single, wide, opaque component, while the dominant component in systems (a) and (b) are strong enough to qualify as \usmgii\ absorbers on their own.  While many intermediate strength (i.e., 1\AA $\la$ \W\ $\la 2$\AA) \mgii\ systems break up into multiple weaker components not resembling those shown in figure~\ref{Fig:UVESspecs}, some are dominated by a single broad opaque component.  For example, c.f.\ figure 2 of Mshar et al.\ (2007), which shows the kinematics of a sample of \mgii\ absorbers ranging from \W$=0.3$\AA\ to \W$=2$\AA, as well as a pair of \usmgii\ systems.  Returning to Q0747+035, could two separate weaker systems be the cause of the apparent \usmgii\ absorber?  The width of the absorption in the SDSS spectrum implies a system dominated by a single opaque component (\S\ref{Sec:kin}).  Thus, this scenario would require two \W$\simeq 2$\AA\ opaque single-component systems at just the right velocity separation ($\delta v \simeq$ 200 \kms).  While possible, we consider this scenario to be less likely than the outflow (starburst-driven or tidally-stripped) scenarios.  The multiple overlapping-absorber scenario is less appealing as an explanation for the velocity spread in the Q1417+011 system due to the extreme REW and that both galaxies have velocities $\approx 700$~\kms\ below the red-edge of the observed absorption. 

\begin{figure}
  \centering
  \includegraphics[angle=0,width=80mm]{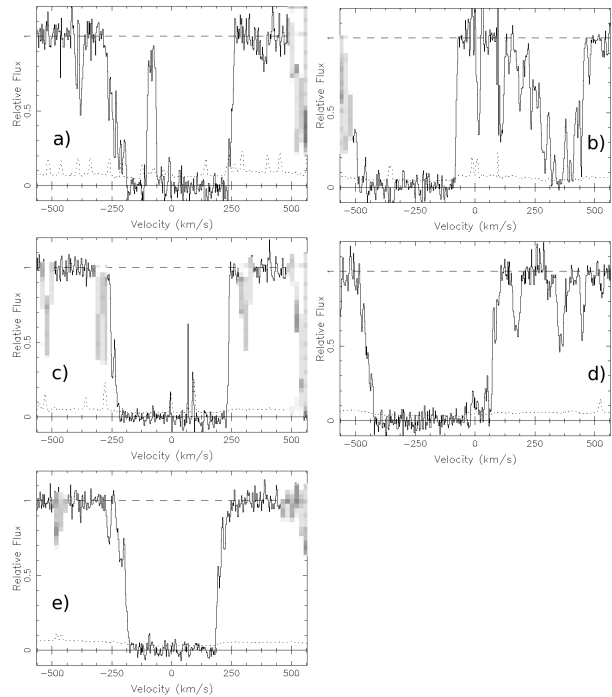}
  \caption{VLT/UVES spectra of $z\simeq 2$ \usmgii\ absorbers (Nestor et al., in preparation).  Panels (a), (c) and (e) show the $\lambda2796$ transition, while panels (b) and (d) show $\lambda2803$.  Blends from other lines are pixelated.  The REW of systems (c), (d) and (e) are dominated by a single, wide, opaque component, while the dominant component in systems (a) and (b) are strong enough to qualify as \usmgii\ absorbers on their own.}
  \label{Fig:UVESspecs}
\end{figure}

Gaseous disks do in some cases contribute to \mgii\ absorption kinematics, the accretion of cool gas must occur at these redshifts and may lead to absorption in \mgii, and some close pairs of \mgii\ systems must exits.  The magnitude of the velocity extent of the {\it strongest} \mgii\ absorbers, however, favours outflows.  This conclusion is strengthened by the presence of galaxies hosting recent starbursts in both of the fields in this study.  As we have mentioned above, however, it is not as clear that the starbursts are actually driving the outflows.  Tidally-stripped gas would likely also span a relatively large range in velocity and the interaction would be expected to trigger star formation episodes, making such a scenario difficult to distinguish observationally from a star formation-driven wind.  The presence of two $z \simeq z_{abs}$ galaxies in both of the fields in this study is consistent with this scenario.  Typical velocity spreads in stripped gas might be expected to be of the order of the velocity dispersion of groups, where gas-rich major interactions are most common.  Although the velocity spreads of (relatively) modest \usmgii\ systems are comparable to that of typical groups, the strongest systems, including the system toward Q1417+011, exceed typical group dispersions.  Furthermore, we note that many \usmgii\ systems appear isolated in the NTRQ imaging data.  While those data are of insufficient quality to rule out minor interactions, which may in principle trigger starbursts, no evidence of the major interactions likely needed to strip large amounts of gas across such huge velocity spreads, such as distorted morphologies or tidal tails, is seen in most fields.  Regarding AGN, we lack the necessary emission-line diagnostics to discriminate between starbursts and LINER/Seyfert galaxies.  
However, the 4000\AA\ break and Balmer absorption strengths clearly indicate recent/ongoing starbursts.  Nonetheless, while the data and kinematics favour star formation-driven outflows over gas stripped from galaxies undergoing a high mass-fraction interaction or AGN-driven winds, the latter scenarios should not be dismissed, particularly as an explanation for some fraction of the less-extreme \usmgii\ systems.  Either way, each of these scenarios involve outflowing gas associated with star-forming galaxies; the primary difference being the energy source responsible for the kinematics.  

\section{Implications} 
This work presents the first two \usmgii\ system host galaxies studied in detail selected from a statistically-understood quasar absorption line sample.  We have shown that outflows driven by galactic star formation appear to be the source of the strongest \mgii\ absorbers (see also Rubin et al.\ 2009; Steidel et al.\ 2010).  Significantly, these systems are identified {\it without direct dependence on galaxy luminosity.}  Thus, \usmgii\ systems provide a method of tracing galactic winds -- and thus high densities of star formation in the Universe  -- selected in a manner complementary to, and without the biases of, emission-based methods.  

We have now collected imaging data for the majority of relatively low-redshift \usmgii\ systems in the SDSS spectra through the fourth data release and intend to investigate the nature of the host galaxies using the entire sample.  Such larger studies are needed to accurately map the connection between \W\ and the detailed properties of the associated galaxies, which will provide important constraints on our understanding of the global evolution of star-forming galaxies.  In this section, we discuss some of the implications of this connection.

\subsection{Strong \mgii\ absorbers}
We have argued that the presence of the {\it strongest} \mgii\ absorbers is related to highly enhanced levels of star formation in galaxies.  However, the connection between star formation and \W\ discussed by M\'{e}nard et al.\ (2009) is significant for \mgii\ systems ranging from \usmgii\ absorbers down to at least \W $=0.7$\AA.  USMgII absorber galaxies almost certainly have significant cross section for absorption at all \W$<3$\AA, and therefore are only observed as USMgII systems, as opposed to weaker systems, by the chance alignment of the sightline through the region of larger \W.  (For the same reason, some fraction of systems observed as weaker \mgii\ absorbers must also have significant USMgII absorption cross section.)  Furthermore, the intrinsic \W-distribution exhibits the form of a single exponential above \W $\simeq 0.3$\AA\ that is featureless despite excellent sampling (Nestor et al., 2005; Nestor et al., in prep), suggesting a common underlying cause for \mgii\ absorption at these strengths.  Thus, it may be that {\it all} \W $\ga 0.3$\AA\ \mgii\ systems are associated with galaxies that have experienced enhanced star formation within their past few Gyr.

\subsection{DLAs}
Strong \mgii\ absorbers (\W $\ga 0.3$\AA) are known to trace high-column densities of neutral hydrogen ($N_{HI} \ga 10^{18}$~cm$^{-2}$).  Damped Ly$\alpha$ absorbers (DLAs, defined as having $N_{HI} \ge 10^{20.3}$~cm$^{-2}$), which have near-unity neutral fractions of atomic hydrogen and are generally metal-poor, have long been considered the reservoirs of cool gas for star formation.  Below $N_{HI} \approx 10^{20.3}$~cm$^{-2}$, subDLAs exhibit a significant ionisation fraction and higher metallicities.  Rao, Turnshek, \& Nestor (2006) investigated the relationship between strong \W\ and $N_{HI}$ over $0.2 \la z \la 1.7$, finding: (i) at a given \W\, absorbers span $\simeq 3$ to 4 orders of magnitude in $N_{HI}$; (ii) $N_{HI}$ is not a good predictor of \W, although the lowest $N_{HI}$ subDLAs are predominantly \W$<1$\AA; and (iii) the likelihood of a \mgii\ absorber having $N_{HI} \ge 10^{20.3}$~cm$^{-2}$ correlates with \W.  These trends fit the outflow scenario very well.  The bulk of the \mbox{H\,{\sc i}} in DLAs is almost certainly associated with only a portion of the velocity profile in the corresponding \mgii\ absorption.  Thus, \usmgii\ absorbers and DLAs are largely not arising in the same gas.  As high sSFRs galaxies should also be gas-rich, a sightline passing through a star formation-driven wind may also pass through a gas-rich portion of the galaxy.  Furthermore, that most DLAs are not \usmgii\ absorbers is consistent with the fact that not all gas-rich galaxies having sufficiently high sSFRs to drive outflows.

\subsection{Contribution to the global SFR density}
The relative contribution of \usmgii\ absorber galaxies to the global star formation density at any redshift can be used to infer the proportion of star formation that takes place at high enough star formation surface density to drive winds at that epoch.  We can write this fraction as
\begin{eqnarray}
 f  &=&  \rho^{USMgII}_{SFR} / \rho_{SFR}\nonumber\\
  &=& \left<\mathrm{SFR}\right> \times \frac{dn}{dV} / \rho_{SFR}\nonumber\\
 & = & \left<\mathrm{SFR}\right> \times \frac{dn}{dz} \times \frac{dl}{dz}^{-1} \times \left<\sigma\right>^{-1} / \rho_{SFR}
\end{eqnarray}
where $\left<\mathrm{SFR}\right>$ is the average SFR in galaxies selected by \usmgii\ absorption, $dn/dz$ is the line-of-sight number density of \usmgii\ absorbers, $dl/dz$ is the differential proper distance, $\left<\sigma\right>$ is the average cross section for  \usmgii\ absorption, and $\rho_{SFR}$ is the global SFR density.  At $z \sim 0.7$,  $\rho_{SFR} \simeq 0.07$~M$_{\sun}$~yr$^{-1}$~Mpc$^{-3}$ (Hopkins \& Beacom, 2006).  We have an accurate measurement of $dn/dz$ from our SDSS catalogs (Nestor et al., 2005; Nestor et al., in prep.), log($dn/dz$) $= -2.11 \pm 0.04$.  Thus, for given values of the average SFR for galaxies selected by \usmgii\ absorption and the average absorption cross section,
we can determine $f$.  Figure \ref{Fig:rho} shows the values of $\left<\mathrm{SFR}\right>$  and $\left<\sigma\right>$ that result in various values for this fraction.  We note that this fraction represents all galaxies having non-zero cross-section for USMgII absorption, regardless of whether it is actually observed as an USMgII system.
\begin{figure}
  \centering
  \includegraphics[angle=0,width=80mm]{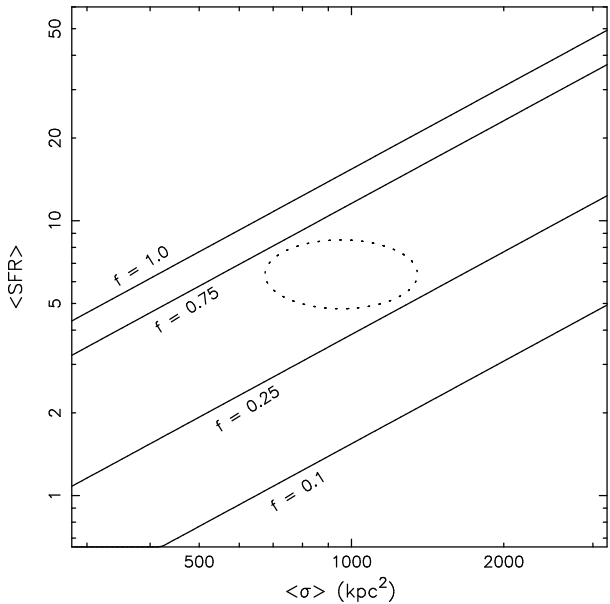}
  \caption{The fraction of the global SF density contributed by \usmgii\ galaxies at $z\sim0.7$, for values of the average SFR in galaxies associated with \usmgii\ absorbers and the average absorption cross section for \usmgii\ absorption.  The ellipse represents the range preferred by the data in conjunction with our model, though the actual uncertainties are unclear (see text).}
  \label{Fig:rho}
\end{figure}

The results of M\'{e}nard et al.\ allow us to accurately compute the average \oii\ luminosity from \usmgii\ galaxies emitted at low-enough $b$ to the quasar sightline to fall within the SDSS spectroscopic fiber used to obtain the quasar spectrum.  This leads to a very firm lower-limit of $\left<\mathrm{SFR}\right> > 0.24$~M$_{\sun}$yr$^{-1}$.  Using the images presented in NTRQ, we determine the fraction of galactic emission, on average, falling within the fiber to be $\sim 30$ per cent, or $\sim 17$ per cent when weighting by luminosity.  Finally, we determine the average reddening corrections for \oii\ from Moustakas, Kennicutt, \& Tremonti (2006) considering the typical luminosity of \usmgii\ galaxies in NTRQ to be a factor of $\sim 6$.  Thus, we estimate $\left<\mathrm{SFR}\right> \sim 5$ to 8 ~M$_{\sun}$yr$^{-1}$.  This value is consistent with the uncorrected SFRs presented in this work.  While it is lower than our inferred dust-corrected SFRs, we again note that the two systems in this study correspond to two of the brightest, highest-$b$ host galaxies from NTRQ, and therefore may be expected to have SFRs exceeding the average for \usmgii\ systems in general.  To approximate $\left<\sigma\right>$, we consider the distribution of impact parameters in NTRQ and a bi-conic model geometry with a range of opening angles and base-widths, and 
account for non-uniformity of the absorption covering factor.\footnote{The increase in the $b$-distribution with $b$ to $\sim$ 15 or 20\,kpc in the NTRQ sample is consistent with the average absorption covering factor being $\sim$ constant with radius out to this distance.  At larger radius, the covering factor must drop -- but remain non-zero -- out to $\sim$ 60\,kpc, to account for the $\sim$ 50 or 30 per cent of absorbers spread over larger values of $b$.}  Doing so, we favour values for $\left<\sigma\right>$ of $\approx 675$ - 1350\,kpc$^2$ (c.f.\ the approximation of $\left<\sigma\right> \simeq 300$\,kpc$^2 $ in Nestor et al., 2005).  We show these ranges of $\left<\mathrm{SFR}\right>$ and $\left<\sigma\right>$ as the ellipse in figure \ref{Fig:rho}.  The preferred range for $f$ is then 0.25 to 0.75, indicating a significant contribution to the global SFR at $z\simeq 0.7$ from \usmgii\ selected galaxies.

We caution that the above calculation for the preferred range of $f$ relies on untested assumptions about the absorption geometry and informal estimation of the uncertainty in the average SFR.   Thus, it should be considered suggestive until future work is able to better constrain these quantities.  Nonetheless, the results have interesting implications on the nature of star formation as a function of redshift.  If $f$ is indeed close to the global value at a given redshift, it implies that the bulk of the star formation at that epoch takes place at high enough star formation surface density to drive winds.  Alternatively, epochs where $f$ is found to be only a small fraction of the global star formation density, $\rho_{SFR}$ must be dominated by relatively diffuse star formation.  Finally, if it were to be found that $f > 1$ at any epoch, it would suggest a significant contribution from faint dwarf galaxies which are being unaccounted for in the current \mbox{H$\alpha$}} or UV surveys yet have high enough SFR surface densities to drive winds.

\section{Summary}
We have presented a deep imaging and spectroscopic study of the fields of two systems from the NTRQ sample of \usmgii\ systems.  In each field we find that there are two galaxies at the absorption redshift having strong emission lines of \oii, \oiii, and \hb.  The emission line fluxes indicate the galaxies are metal rich and have significant ongoing star formation.  We employed SED template fitting to estimate stellar masses, which indicates relatively high \mbox{sSFRs}.  Analysis of the 4000\AA\ break and Balmer absorption strengths indicate the $z\simeq z_{abs}$ galaxies toward Q0747+035 underwent a starburst $\sim$~1 Gyr in the past, while those towards Q1417+001 are currently in a starburst phase.  It is extremely unlikely to find galaxies with such properties at the same location as the \usmgii\ absorption unless they are related in some way to the low-ion velocity spreads that define \usmgii\ systems.  We consider various popular models for the nature of \mgii\ absorption systems.  Given the (post-)starburst natures of the galaxies, their velocities relative to the observed absorption kinematics, the burst ages, together with the impact parameters of the sightlines to the galaxies, we conclude that starburst-driven galactic winds are the most likely causes of the \usmgii\ absorption.  However, a scenario in which gas is tidally stripped by galaxy-galaxy interactions which simultaneously trigger starbursts is also consistent with the data.  While past studies have found blueshifted absorption in the spectra of galaxies at cosmological distances, identifying outflows in this manner unambiguously demonstrates that the material reaches the IGM.\footnote{Also see Steidel et al.\ (2010).}  Finally, we estimate that the star formation density traced by \usmgii\ absorbers is, roughly, within at least an order of magnitude of the total global density, indicating that, though rare, \usmgii\ absorbers are a powerful tracer of star formation in the Universe.

\section*{Acknowledgments}
DBN and BDJ acknowledge support from the STFC-funded Galaxy Formation and Evolution programme at the Institute of Astronomy.  
VW acknowledges European Union support from a Marie Curie
Intra-European fellowship.  Based on observations obtained at the Gemini Observatory, which is operated by the
Association of Universities for Research in Astronomy, Inc., under a cooperative agreement
with the NSF on behalf of the Gemini partnership: the National Science Foundation (United
States), the Science and Technology Facilities Council (United Kingdom), the
National Research Council (Canada), CONICYT (Chile), the Australian Research Council
(Australia), Minist\'{e}rio da Ci\^{e}ncia e Tecnologia (Brazil) 
and Ministerio de Ciencia, Tecnolog\'{i}a e Innovaci\'{o}n Productiva  (Argentina).

\bsp

\label{lastpage}

\end{document}